\documentclass[preprint]{iucr}

\journalcode{J}              
\usepackage{xcolor}
\usepackage{graphicx}
\usepackage{amssymb}
\usepackage{amsmath}
\usepackage{tabularx}
\usepackage{float}
\usepackage{enumitem}
\usepackage{wrapfig}
\usepackage{comment}

\DeclareMathOperator{\Res}{Res}

\begin{document}                   

\title{Measuring the Burgers Vector of Dislocations with Dark-Field X-ray Microscopy}

\author[a,b,c]{Dayeeta}{Pal}
\author[a,b,c]{Yifan}{Wang}
\author[d]{Ramya}{Gurunathan}

\cauthor[a,b,c]{Leora}{Dresselhaus-Marais} 
{leoradm@stanford.edu}

\aff[a]{Stanford University, Department of Materials Science \& Engineering, Stanford, CA, \country{USA}}
\aff[b]{SLAC National Accelerator Laboratory, Menlo Park, CA, \country{USA}}
\aff[c]{SIMES, Menlo Park, CA, \country{USA}}
\aff[d]{National Institute of Standards and Technology, Gaithersburg,MD \\}

\maketitle                        

\begin{synopsis}
This work introduces the first direct comparison of contrast arising from dislocations with Dark-Field X-Ray Microscopy (DFXM) and the established Transmission Electron Microscopy (TEM). Following that, we propose a method to directly measure the burgers vector of a dislocation based on the symmetry with which DFXM scans across a rocking curve for a given diffraction peak.
\end{synopsis}

\begin{abstract}
The behavior of dislocations is essential to understand material properties, but their subsurface dynamics that are representative of bulk phenomena cannot be resolved by conventional transmission electron microscopy (TEM). Dark field X-ray microscope (DFXM) was recently demonstrated to image hierarchical structures of bulk dislocations by imaging lattice distortions along the transmitted X-ray diffracted beam using an objective lens. While today’s DFXM can effectively map the line vector of dislocations, it still cannot quantify the Burgers vector required to understand dislocation interactions, structures, and energies. Our study formulates a theoretical model of how DFXM images collected along specific scans can be used to directly measure the Burgers vector of a dislocation. By revisiting the “invisibility criteria” from TEM theory, we re-solve this formalism for DFXM and extend it to the geometric-optics model developed for DFXM to evaluate how the images acquired from different scans about a single \{hkl\} diffraction peak encode the Burgers vector within them. We demonstrate this for edge, screw, and mixed dislocations and discuss the observed symmetries. This work advances our understanding of DFXM to establish its capabilities to connect bulk experiments to dislocation theory and mechanics.
\end{abstract}

\section{Introduction}

Dislocations are essential to materials science because these atomic-scale linear defects in crystals govern the macroscopic material properties. A dislocation is a linear defect arising from a truncated plane of the lattice generated by localized strain fields. Dislocations store deformation energy inside a lattice, but are notoriously difficult to observe experimentally -- especially in the statistically significant numbers relevant to bulk properties \cite{kuhlmann1999theory}. The properties arising from dislocations are related to the complex networks of dislocation patterns that form under different loading conditions. The relevant forces and interactions of dislocations span several length scales, with an \AA-scale core that produces large-range, micron-scale displacement fields.

Several different approaches are used to measure dislocations in metals. Atomic Force Microscopy (AFM) analyzes the surface roughness at the atomic scale, resolving atomic steps in the surface arising from dislocation slip \cite{chen2020atomic}. More recently, high resolution Scanning Electron Microscope (SEM) has used digital image correlation algorithms to study dislocation structures based on similar surface steps to understand the localization of plasticity \cite{stinville2016sub}. Transmission Electron Microscopy (TEM) map dislocations via their displacement fields, by imaging along the beam diffracted by the lattice (a.k.a. diffraction contrast) \cite{taheri2016current}. 
Electron and force microscopy have imaged dislocations since 1956 \cite{hirsch1956lxviii} in thin samples or at material surfaces, but we still struggle to understand the analogous phenomena deep beneath the surface. 

TEM has pioneered the use of diffraction contrast imaging to observe dislocations via dark-field electron microscopy (a.k.a. \textit{diffraction contrast} TEM).
That work demonstrated that by imaging selected diffraction conditions, the displacement fields surrounding a dislocation either attenuate the image intensity (a.k.a. the \textit{strong-beam condition}) or enhance the image signal.
This change in signal arising from local displacement fields defines the \textit{contrast mechanism} governing dark-field or diffraction based imaging in TEM and X-ray topography. 

As such, diffraction-contrast images are produced as a result of displacement field, not just the dislocation core, which perturb the diffraction condition for electrons and X-rays. 
By rocking the crystalline sample, the diffraction condition is altered due to the tilt of the incident beam with respect to the orientation of the crystal planes, causing different regions surrounding the dislocations to be observed. 

To extend TEM's insights into dislocations beneath a crystal's surface, X-Ray topography (XRT) emerged in 1958 \cite{lang1958direct} as an X-ray analogue of dark-field TEM. XRT collects a near-field image along the X-ray diffracted beam, resolving signal transmitted through thick single-crystalline samples. However, XRT has always been limited in its resolution and interpretability beyond low dislocation density systems (e.g. diamond, silicon) by the near-field imaging geometry (analogous to radiography) \cite{danilewsky2020x}. 
With significant upgrades in the coherence and flux of synchrotron radiation, other techniques have become able to resolve subsurface dislocations such as Bragg ptychography \cite{hammarberg2023strain} and Bragg Coherent Diffraction Imaging (BCDI) \cite{clark2015three}. These techniques can image 3D dislocation structures at up to 10-nm resolution, but lack the ability to time-resolve those dynamics due to their requisite scans to enable phase reconstruction. 
  
First published in 2015 \cite{simons2015dark}, dark-field X-ray microscopy (DFXM) has recently emerged as a novel technique to directly image subsurface dislocations and their dynamics at nm-$\mu$m scales \cite{jakobsen2019mapping,DresselhausMarais2021}. 
Analogous to dark-field TEM and XRT, DFXM captures images along the X-ray diffracted beam -- using an X-ray objective lens to magnify the images and refine the contrast mechanism. 
In particular, the low numerical aperture of the objective acts as a ``reciprocal-space filter,'' offering direct access to distinguish between axial and shear strain fields in the crystal \cite{Poulsen2017}.
The recent development of a forward model for DFXM using geometrical optics provides an opportunity to evaluate how to manipulate the contrast mechanism to uniquely understand dislocations \cite{Poulsen2021}. Their study focused on the single-dislocation level, and more recent work has explored dislocation structures comparing the geometric vs wave optics approaches \cite{borgi2024simulations}. More recently, our group has extended this study to large dislocation networks by integrating molecular dynamics and discrete dislocation dynamics into the geometric optics model (Wang et al., in Preparation). 

Full dislocation characterization requires measurement of the line vector that defines the core, and the Burgers vector that defines the direction and magnitude of shear it imparts onto the lattice. 
While 3D DFXM was recently able to resolve the line vectors \cite{yildirim2023extensive}, interpretation of the Burgers vector has always required assumptions. At this time, direct measurement of a dislocation's Burgers vector with DFXM has not been demonstrated.

In this work, we explore the characterization of the Burgers vector of dislocations in face-centered (FCC) materials using the DFXM forward model. 
We begin by introducing how TEM measures the Burgers vector through the ``invisibility criterion'' on different \{hkl\} peaks. 
To understand if this invisibility criterion is also applicable to DFXM, we describe the DFXM geometric optics model \cite{Poulsen2021}. We formulate a simplified model for DFXM's sensitivity to different deformation gradient fields that could contribute to signal from dislocations.
We then go on to discuss how DFXM would sample a dislocation by demonstrating the different imaging conditions from points along a 2D rocking curve scan.
From those results, we confirm that the TEM approach to measuring dislocation visibility based on a $\vec{g}\cdot \vec{b}$ criterion is valid for DFXM, but only in specific imaging conditions with long setup times.
To formulate a faster approach to Burgers vector measurement, we derive the invisibility criterion for DFXM for a single \{hkl\} peak for a simplified system. We observe how for a realistic single \{hkl\} experiment the 2D rocking curve can allow one to directly measure the Burgers vector, which we confirm in our simulations.
Finally, we illustrate how the symmetry native to edge dislocations in FCC crystals generates complementary imaging signals in our DFXM images, and conclude by extending our edge dislocation simulations to screw, and mixed dislocations.
Our work illustrates a key step forward in using DFXM to measure dislocation behaviors relevant to mechanical properties in metals.

\section{TEM theory of burgers vector identification}\label{sec: TEM theory of burgers vector identification}

We begin by reviewing the TEM theory that has been used to measure the Burgers vector, $\Vec{b}$, using dark-field contrast. In this section we introduce the first theory of electron diffraction and the Howie-Whelan equations describing how the projection of the displacement field surrounding a dislocation onto the diffraction vector $\Vec{g}$ \footnote{TEM literature uses the notation $\vec{g}$ for diffraction vector while the DFXM community uses $\vec{Q}$} gives rise to contrast. We conclude by explaining how this leads to the invisibility criterion that is used to measure $\vec{b}$ with TEM. 

Howie et al. \cite{head2012computed} formulated a wave-optics theory based on dynamical diffraction to describe the intensity of each point on a dark-field TEM micrograph. The solution to the Howie-Whelan equations for the single-crystal diffraction condition in a pristine crystal reveals how the electron wavefunction oscillates sinusoidally between the direct beam and the diffraction vector. The resulting Howie-Whelan equations describe the conservation of momentum and energy using the Takagi Toupin formalism and ended up with a complex set of differential equations to describe the full visibility of dislocations. They encapsulate the intricate interplay between momentum and energy conservation in determining the how the dislocation observed. 

Howie simplified their equations to describe whether a dislocation would be visible or invisible (a.k.a. the \textit{invisibility criteria} to the expression 
\begin{equation}
    \beta^{'} = \frac{d\left(\vec{g} \cdot \vec{R}\right)}{dZ},
\end{equation}
where $\beta^{'}$ is the displacement gradient field\footnote{$\frac{\partial \Vec{u}}{\partial x}$ in DFXM literature}, $\vec{R}$ is the atomic displacement field caused by the dislocation\footnote{$\vec{u}$ in DFXM literature} and $Z$ is the depth being observed within the sample\footnote{$x$ in DFXM literature}. When $\beta^{'}=0$, then 

\begin{equation} \label{eq:gRZero}
    (\vec{g} \cdot \vec{R}) = 0,
\end{equation}
producing an image that is identical to one obtained from a pristine crystal. For this condition, we can say the dislocation becomes invisible.

By substituting the variable $\vec{R}$ for the expression from mechanics, it can be shown that Equation~\ref{eq:gRZero} gives the form
\begin{equation}
    \vec{g} \cdot 
    \bigg[\vec{b}\theta + \vec{b}_{e} \bigg\{\bigg(\frac{\sin 2\theta}{4\left(1 - \nu\right)}\bigg) 
    + (\vec{b}\times\vec{\ell})\bigg\{\bigg(\frac{\big(1 - 2\nu\big)}{2\big(\vec{\ell} -\nu\big)}\bigg)\ln{\Vec{r}} 
    + \frac{\cos 2\theta}{4(\vec{\ell} -\nu)}\bigg\}\bigg] = 0,
\end{equation}
which simplifies to
\begin{equation} \label{eq:fullform}
    \Vec{g} \cdot \Vec{b}\theta + \Vec{g} \cdot b_{e} {\left(\frac{\sin 2\theta}{4\left(1 - \nu\right)}\right)} + g \cdot (\vec{b}\times \vec{\ell})\{\left(\frac{\left(1 - 2\nu\right)}{2\left(\vec{\ell} -\nu\right)}\right){\ln} \vec{r} + \frac{\cos 2\theta}{4\left(\vec{\ell} -\nu\right)}\} = 0.
\end{equation}
Since Equation \ref{eq:fullform} is a summation of three terms, it may be approximated that invisibility will occur when each term individually equates to 0. For the full treatment of dislocation invisibility, we include a derivation of the most general form taken directly from $\Vec{g} \cdot \Vec{R} = 0$ in Appendix A. 
With their approximation, the TEM invisibility criterion used today may be derived as $\vec{g} \cdot \vec{b}=0$ for screw dislocations, which have the minimum number of out-of-plane components in their displacement field. For edge dislocations, an additional $\vec{g} \cdot (\vec{b}\times\vec{\ell})=0$ must also be satisfied due to their maximum number of out-of-plane components. 

\begin{figure}
\centering
\includegraphics[width=0.5\textwidth]{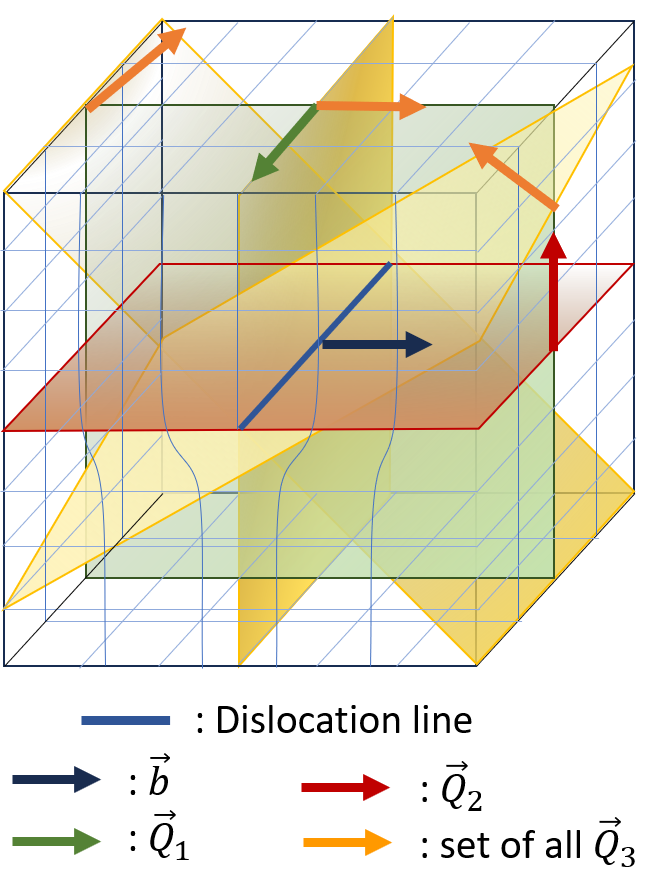}
\caption{Schematic depicting a lattice with an edge dislocation (blue line). Images along the green and red planes produce an invisible dislocation, while images collected along the yellow/orange planes show dislocation visibility.}
\label{fig:1_diffraction_vectors}
\end{figure}

 In a TEM experiment, one collects images along several diffraction peaks and identifies the two mutually orthogonal diffraction peaks that cause the dislocation to become invisible. As shown in Fig.~\ref{fig:1_diffraction_vectors} the two $\vec{Q}$s that generate invisibility uniquely define the planes orthogonal to $\vec{b}$. Using the notation from the figure, we can express this mathematically as
\begin{align}
    \vec{Q}_1\cdot\vec{b} =& 0 \\
    \vec{Q}_2\cdot\vec{b} =& 0 \\
    \vec{Q}_3\cdot\vec{b} \neq& 0. 
\end{align}
Taking these three results together, we can compute that the Burgers vector based on the cross product of the vectors for which the dislocation was invisible,
\begin{equation}
    \vec{b} =  \vec{Q_1} \times \vec{Q_2}.
\end{equation}
In practice, this measurement is rarely done as imaging along only three diffraction peaks, but instead as a cascade of measurements along many diffraction planes. The result is that the Burgers vector for a full population of dislocations may be observed based on the two invisibility vectors for each dislocation.

\section{Review of the DFXM Forward Model} \label{sec:Defining DFXM sensitivity with the Forward Model}

To connect the TEM theory derived above to DFXM experiments, we spend this section introducing the formalism for our DFXM simulation model. We focus for this study on a simplified geometric optics formalism demonstrated previously \cite{Poulsen2021}. While this differs from the wave optics approach from the Howie-Whelan equations, we do include results from the wave optics formalism derived elsewhere for qualitative validation of the similarity between models, as has been done previously \cite{borgi2024simulations}.
 
We then conclude by simulating a simplified description of the sensitivity of the shear displacement fields and how they contribute to the signal observed by DFXM.  

In this work, we use the geometrical optics formulation for the DFXM forward model that is derived by Poulsen \textit{et al.} in \cite{Poulsen2017, Poulsen2018, Poulsen2021}. The 2017 work derives analytical expressions for the optical components in DFXM using a ray transfer matrix approach \cite{Poulsen2017}. The 2018 work then expands upon the previous study by extending the optical equations into a resolution function that defines the reciprocal space information \cite{Poulsen2018}. The 2021 paper then brings these two works together, adding a micromechanical model to describe diffraction in order to compile a full DFXM forward model to simulate DFXM images \cite{Poulsen2021}. Their work ends with simulations of dislocations based on the deformation gradient tensor, $\textbf{F}=\nabla \textbf{u} + \textbf{I}$. The simulations have been performed using Python code in the link https://github.com/leoradm/DFXM-BurgersVector.

\begin{figure}
  \centering
  \includegraphics[width=\linewidth]{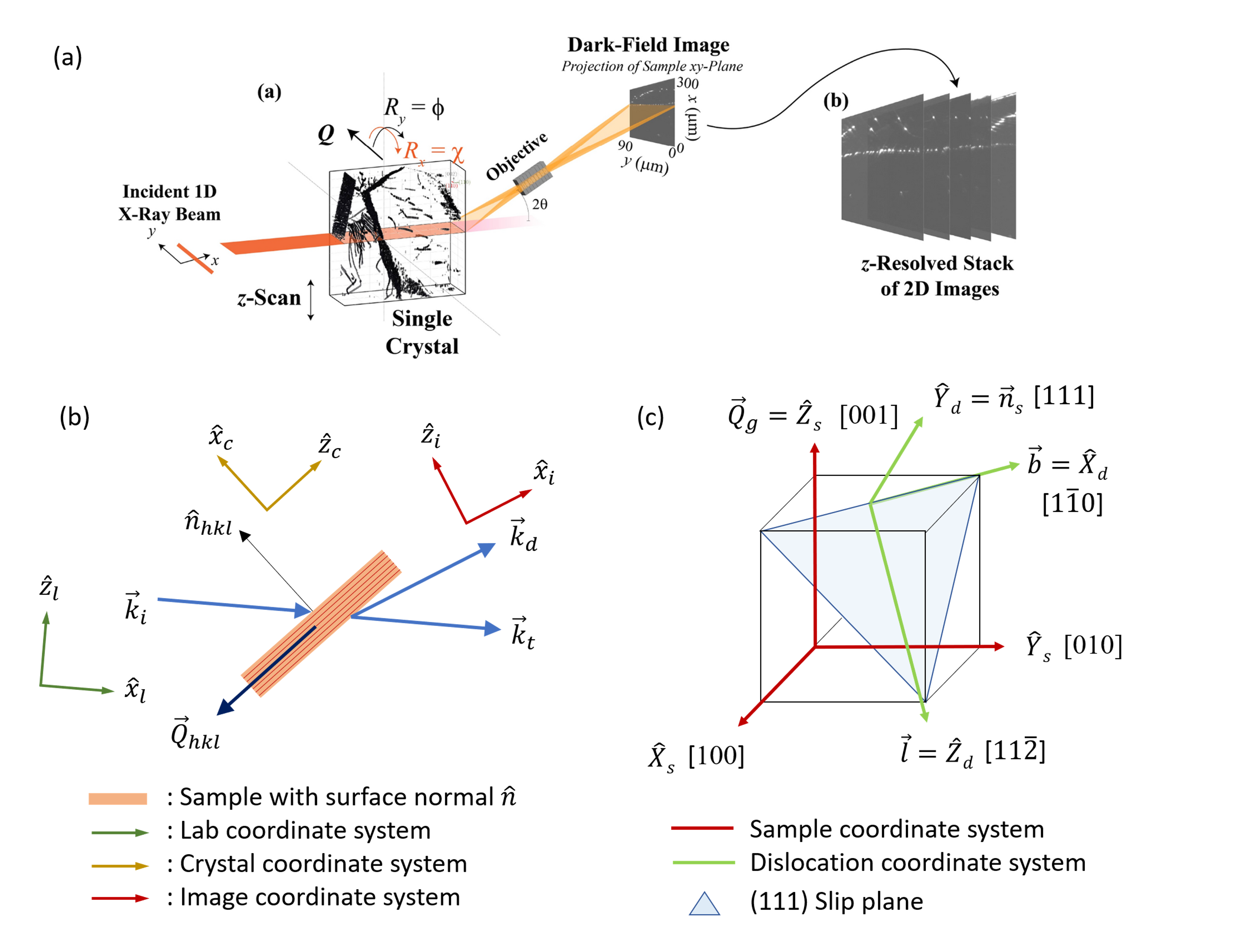}
  \caption{(a) Dark Field X-Ray Microscope \cite{yildirim2023extensive}. (b) Schematic of the five different coordinate systems defined for a DFXM setup. We include the sample, crystal planes, and incident/diffracted beams for reference to guide the definition of each coordinate system, as described in the text. (c) An edge dislocation defined by $\Vec{b}$, $\Vec{\ell}$ and $\Vec{n}$ in an FCC lattice and its relation to the sample coordinate system \cite{Poulsen2021}}
  \label{fig:DFXM}
\end{figure}

For mathematical convenience, this formalism defines a set of five related coordinate systems that define the different optical/experimental components. We summarized the equations defining each coordinate system in Table~\ref{tab:coordinate systems}, and illustrate these relationships in Fig.~\ref{fig:DFXM}b. The \textit{laboratory system} is defined by the incident X-ray beam, while the \textit{crystal system} is defined by the sample's surface normal. The sample system is then defined by the diffraction vector, $Q_{hkl}$, being measured for this DFXM experiment, while the \textit{grain system} is defined based on the specific domain of the crystal under consideration. For single crystals, the grain and sample systems are equivalent, but they are distinguished for convenience in polycrystalline systems. The imaging system is used to defined the optical components following the crystal, and is thus aligned to the X-ray diffracted beam. Finally, the \textit{dislocation system} is defined based on the Burgers and line vectors of the dislocations for continuum elastic models. 

\begin{table} 
\centering 
\renewcommand{\arraystretch}{1.3}
\begin{tabular}{|c| c| c| c|}
    \hline 
    \textbf{Lab System} & \textbf{Sample System} & \textbf{Crystal System} & \textbf{Imaging System} \\[0.5ex] 
    \hline 
    $\hat{k}_i = \hat{x}_{\ell}$ & $\hat{Q}_{hkl}=\hat{x}_s$ & $\hat{n}_{hkl}=\hat{x}_c$ & $\hat{k}_d=\hat{x}_i$ \\ 
    \hline 
    \hline
\end{tabular}
\caption{Coordinate systems to describe DFXM, as described in the text and in \cite{Poulsen2021}.}
\label{tab:coordinate systems}
\end{table}

For mathematical convenience, we define the dislocation coordinate system for an edge dislocation as $\hat{x}_d$, $\hat{y}_d$ and $\hat{z}_d$ are unit vectors in the direction of $\Vec{b}_s$ = $1/2[1\Bar{1}0]_s$,  $\Vec{n}_s$ = $[11\Bar{1}]_s$, and $\Vec{\ell}$ = $[112]_s$ respectively. Therefore, the rotation matrix that converts a vector in the dislocation coordinates $\Vec{r}_{d}$ into a vector in the sample (grain) coordinates $\Vec{r}_{g}$ is 
\begin{equation} \label{}
\Vec{r}_{g}=\mathbf{U}_{d}\Vec{r}_{d},
\end{equation}
where 
\begin{equation} \mathbf{U}_{d}=\begin{bmatrix}
b_{x,g} & n_{x,g} & \xi_{x,g} \\
b_{y,g} & n_{y,g} & \xi_{y,g} \\
b_{z,g} & n_{z,g} & \xi_{z,g}
\end{bmatrix}.
\end{equation}

As such, the dislocation system changes with the grain system, as the sample is rotated. Following the procedure mentioned in Table~\ref{tab:coordinate systems}, the goniometer rotations alter the relationship between the lab and sample systems by reorienting $\Vec{Q}$, without changing the relationships between the grain and dislocation systems. A $\phi$ or $\chi$ scan with DFXM would thus transform the $\Gamma$ matrix required to convert the lab to sample system, but would rotate both the grain and dislocation systems implicitly because they are defined based on the sample system.

As derived in Poulsen et. al., 2021 $\vec{Q}_g$ is related to Miller indices $(h,k,\ell)$ by
\begin{equation}
    \vec{Q}_g = \mathbf{B}_0 \begin{pmatrix}
           h \\
           k \\
          \ell
     \end{pmatrix}.
\end{equation}
This gives an expression for the undeformed crystal lattice based on the lattice parameters defined within $\textbf{B}$ \cite{Poulsen2021}. The formalism may also  be extended to a deformed crystal lattice by defining $\textbf{B} = (\textbf{F}_g)^{-T} \textbf{B}_0$ to describe the deformed crystal lattice. We thus get an expression for the deformed diffraction vector 
\begin{align}
  \vec{Q}_g =  (\mathbf{F}^g)^{-T} \mathbf{B}_0 
    \begin{pmatrix}
           h \\
           k \\
          \ell
     \end{pmatrix}=\mathbf{B} \begin{pmatrix}
           h \\
           k \\
          \ell
     \end{pmatrix}. \label{eq:Joel}
\end{align}
as described in Poulsen et. al., 2021 and originally derived in \cite{bernier2011far}.

Components of the strain tensor are measured through small changes in $2\theta$, while the local rigid body rotations can be measured by a large rotation of the diffraction spots at a fixed diffraction angle of $2\theta$. The real-space deformation tensor then needs to be mapped to reciprocal space to study the effects of deformation on the diffraction vector $\mathbf{Q}$. To do so, Poulsen et al., defined the inverse deformation tensor as 
\begin{equation}  
\mathbf{H}^{\mathrm{g}} = (\mathbf{F}^{\mathrm{g}})^{-\mathrm{T}} - \mathbf{I}.
\label{eq:inv deformation tensor}
\end{equation}

The reciprocal lattice vector at a given position in the deformed sample ($\mathbf{q}_{\mathrm{g}}$) is then equal to:
\begin{equation}
    \vec{q}_{\mathrm{g}} = \mathbf{H}^{\mathrm{g}}\vec{Q}_0,
\end{equation}
where $\vec{Q}_0$ is the undeformed lattice vector in the grain system.

With the coordinate systems and the conditions of diffraction defined, we focus on how intensity on the detector is defined, as is done in the Howie-Whelan equations above. In this formalism, we define intensity that reaches the detector as the convolution of the diffraction condition and a resolution function. The resolution function $\Res(\vec{r}, \vec{q},y_i', z_i')$ is formally defined as a 6D function describing the vectors that can traverse the optics within the DFXM system. The full function of $\Res(\vec{r}, \vec{q},y_i', z_i')$ is based on the divergence of the incident beam, the aperture of the objective lens (a compound refractive lens, CRL), and the bandwidth of the incident beam, based on Equations 42-45 in \cite{Poulsen2021}, 
\begin{align}
    q_{\mathrm{rock}} 
    & = 
    -\frac{\zeta_v}{2} - (\theta - \theta_0) + \phi 
    \label{eq2}
    \\
    q_{\mathrm{roll}}
    & = 
    -\frac{\zeta_h}{2\sin(\theta)} - \cos(\theta)\eta + \chi
    \label{eq3}
    \\
    q_{\parallel}
    & =
    \frac{\delta E}{E} + \cot(\theta) \left[- \frac{\zeta_v}{2} + (\theta - \theta_0) \right].
    \label{eq4}
\end{align}
In the above equations, we note that the resolution function depends on factors from both the scanning dimensions and the sample setup. As shown in Fig.~\ref{fig:DFXM}, $\phi$ and $\chi$ define the rocking and rolling directions for the crystal at the goniometer, which alter the diffraction condition and shift the span of the resolution function.
We note that the $\chi$ dependence of $q_{\mathrm{roll}}$ was omitted from Poulsen, et al., but is included here for completeness. 
The values $\theta-\theta_0$ define the third direction often scanned during DFXM imaging, referred to as $\Delta \theta$. Experiments typically collect images while rocking along $\Delta \theta$ to probe axial strain along the selected $(hkl)$ plane, while they collect images during scans along $\phi$ and $\chi$ to resolve the shear strains. For polycrystalline samples, images collected during scans along the $\eta$ positions probe the orientational spread of the grains (a.k.a. the mosaicity). 

In Poulsen, et al. 2021, a Monte-Carlo simulation of rays within the horizontal and vertical divergence, $\zeta_h$ and $\zeta_v$, respectively, and the bandwidth, $\delta E / E$, are input the above equations and the results are fitted to Gaussian expressions to formulate $\Res(\vec{q})$.
The micromechanical model uses the displacement gradient field, $\nabla \textbf{u}$, as used in the Howie-Whelan equations to define the diffraction condition using a linear algebra formalism. In the DFXM formalism, $\nabla \textbf{u}$ is reformulated into the deformation gradient tensor field, $\textbf{F}$, to explicitly define the formulas for the fields surrounding dislocation cores. 

Using the above formalisms, the \textit{intensity} on an image may be defined using the six-dimensional integration of
\begin{equation} \label{eq:dfxm-image-intensity}
    I(y_i', z_i') = \iiint_{\vec{r}} \iiint_{\vec{q}}  I_d(\vec{r},\vec{q}) \Res(\vec{r}, \vec{q},y_i', z_i') \,\mathrm{d}^3 \vec{r} \,\mathrm{d}^3 \vec{q},
\end{equation}
to describe a selected point on the detector in the imaging system, $(y_i', z_i')$. In this work, we only consider single crystalline samples, oriented in the vertical diffraction condition, i.e., $\eta = 0$.

Here we now explain how each component of strain or displacement alter the signal in DFXM to quantify the sensitivity of DFXM. 

As can be seen from the above equations, by simply changing the goniometer angles, $\phi$ and $\chi$, we shift the resolution function and thus alter the overlap between $\Res(\vec{r}, \vec{q},y_i', z_i')$ and the reciprocal lattice point of interest for our DFXM experiment.
To propose a strategy to measure the Burgers vector using DFXM, we begin by exploring how DFXM samples a set strain field that includes multiple deformation components.

\section{Defining DFXM sensitivity with the Forward Model}

To begin our study, we define a \textit{Simplified Geometry} that we use in Sections 4-6 to describe an edge dislocation in an fcc aluminum single crystal. Our simplified geometry is referenced against a grain system of $\hat{x}_g=[100]$, $\hat{y}_g=[010]$, $\hat{z}_g=[001]$, and we define the crystal system to be identical to the grain system, i.e.,
\begin{equation}
    \vec{r}_c = \textbf{U} \vec{r}_g
    \hspace{2em} \text{where} \hspace{2em}
    \textbf{U} = \textbf{I}.
\end{equation}

\begin{figure}
  \centering
  \includegraphics[width=\linewidth]{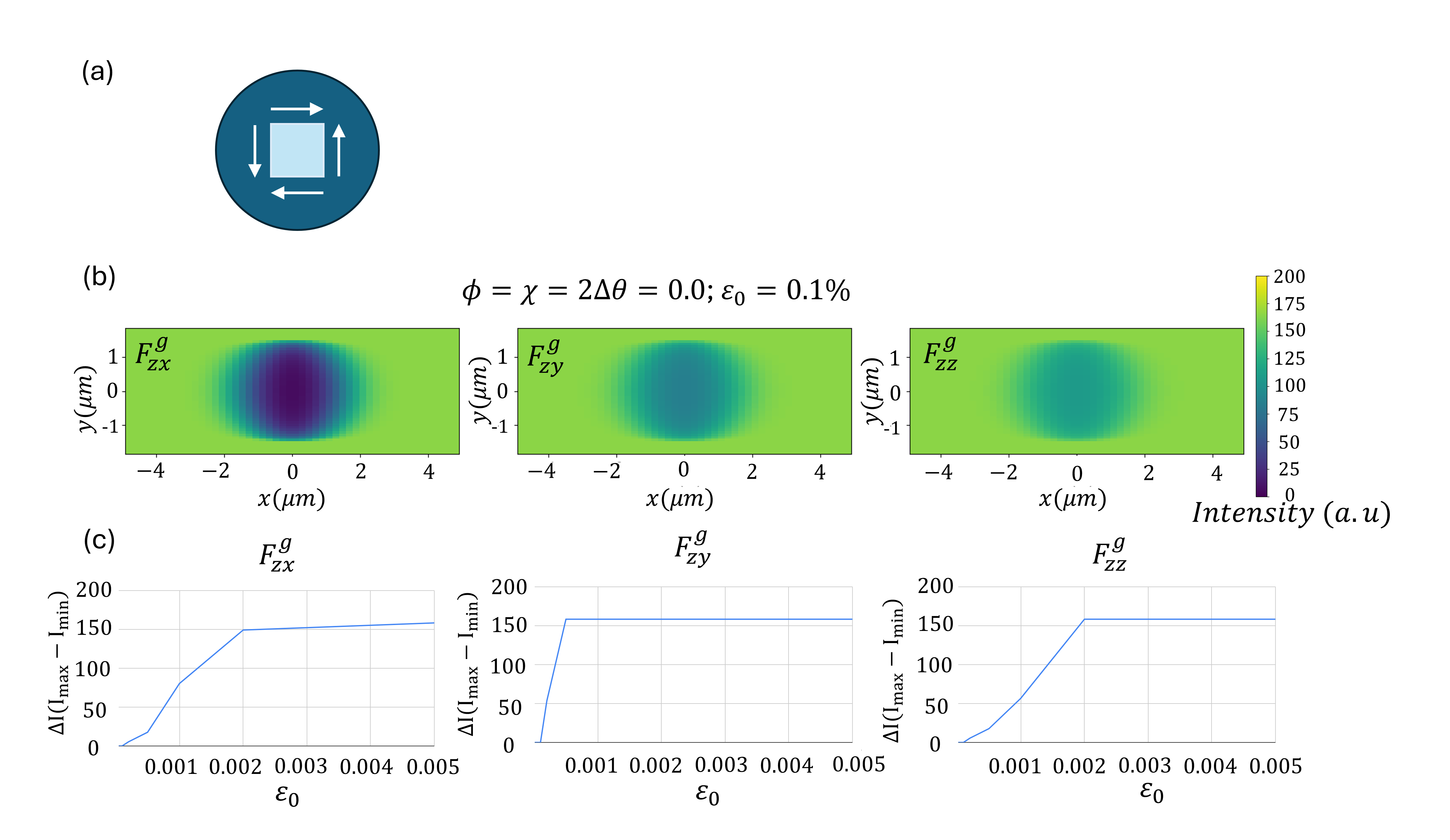}
  \caption{(a) Schematic of the sample configuration experiencing shear. (b) Simulation showing the DFXM image arising from $\varepsilon_0=0.1\%$. (c) The global image contrast (Eq.~\ref{eq:contrast}) for increasing values of $\varepsilon_0$, plotted for $\textbf{F}_{zx}^{(g)}$, $\textbf{F}_{zy}^{(g)}$, and $\textbf{F}_{zz}^{(g)}$.}
  \label{fig:Intensity-strain}
\end{figure}

We define our initial test sample as a pristine crystal whose only deformation field is a sphere of uniform shear that we impose using our forward model. We define our sphere as
\begin{equation}
   r=\sqrt{x^2+y^2+z^2}
    \hspace{2em} \text{for} \hspace{2em} 
    r\leq1.5,
\end{equation}
where $\textbf{F}^{(g)}$ is a piecewise function defined as  
\begin{equation}
    \textbf{F}^{(g)} = \left\{
\begin{array}{ll}
      0 & r> 1.5 \\
      \textbf{F}_{\varepsilon_0} & r \leq 1.5 \\
\end{array} 
\right.
    \hspace{2em} \text{where} \hspace{2em} 
    \textbf{F}_{\varepsilon_0} = \begin{bmatrix}
        \varepsilon_{0} & \varepsilon_{0} & \varepsilon_{0} \\
       \varepsilon_{0} & \varepsilon_{0} & \varepsilon_{0} \\
        \varepsilon_{0} & \varepsilon_{0} & \varepsilon_{0}
    \end{bmatrix}
\end{equation}
for increasing values of $\varepsilon_{0}$ to define the sensitivity of DFXM to each component of $\textbf{F}_{ij}^{(g)}$. A schematic of the sample configuration is shown in Fig.~\ref{fig:Intensity-strain}a, with a sample DFXM image collected for $\phi=\chi=2\Delta\theta=0$, $hkl=[002]$, $2\theta_0=20.73^{\circ}$, $\zeta_v=2.25 \times 10^{-4}$, $\zeta_h=4.2 \times 10^{-4}$ and $\frac{\delta E}{E}=6 \times 10^{-5}$ as shown in Fig.~\ref{fig:Intensity-strain}b.

We demonstrate the trends in DFXM's sensitivity to each component of $\textbf{F}_{ij}^{(g)}$ in Fig.~\ref{fig:Intensity-strain}c. Because we selected the $[002]$ diffraction peak, we note that only these three values of $\textbf{F}_{ij}^{(g)}$ gave non-zero contributions to signal variation in the images. We quantify sensitivity based on the \textit{contrast} we observe in each image, which we define as
\begin{equation} \label{eq:contrast}
    \Delta I = I_{\text{max}} - I_{\text{min}}
\end{equation}
across the entire simulated image, which we plot for $\textbf{F}_{zx}^{(g)}$, $\textbf{F}_{zy}^{(g)}$, and $\textbf{F}_{zz}^{(g)}$ in Fig.~\ref{fig:Intensity-strain}a. 
We note that in this work, we discuss two types of image contrast, the \textit{global contrast}, $\Delta I$ as defined in Equation~\ref{eq:contrast}, and the \textit{local contrast}, $\delta I$ that is defined as the intensity difference between two adjacent pixels. 
In all plots, the increased magnitude of $\varepsilon_0$ creates a corresponding increase in the contrast in DFXM images. This reaches a maximum value of 165.03 that only increases slightly for $\textbf{F}_{zy}^{(g)}$ because of the absence of a vertical aperture. The smallest values of $\varepsilon_0$ that give rise to our maximum image contrast are both $\varepsilon_0=0.002$ for $\textbf{F}_{zy}^{(g)}$ and $\textbf{F}_{zz}^{(g)}$, but is less than half that value for $\textbf{F}_{zx}^{(g)}$ ($\varepsilon_0=0.00035$).

From this simplified model system, we can conclude that $\textbf{F}_{zx}^{(g)}$ is the most sensitive component of $\textbf{F}$ in our DFXM to deformations that contribute to signal in our DFXM images of the crystal in this simplified geometry along [002]. This leads us to the understanding that when observed by DFXM, some of the shear components of the strain field surrounding a dislocation should contribute more strongly than others to image signal. From the simulated images, we can derive a relationship between the strain field of a dislocation line segment and the contrast ($\Delta I$) on the detector using Fig.~\ref{fig:Intensity-strain}b. 

From the plots above, we performed a linear fit to obtain a relation between the DFXM image contrast and strain for $\textbf{F}_{zx}^{(g)}$ as

\begin{equation} \label{eq:deltaI}
    \Delta I = 368336\varepsilon_0 - 14.07.
\end{equation}

Because we formulated our deformation field such that the only image contrast arises from the same $\delta I$ making the global and local contrast equivalent. 

The displacement gradient of an edge dislocation are calculated using Equations \ref{eq:1}-\ref{eq:4}. 

The magnitude of $\textbf{F}_{ij}^g$ around a dislocation is dependent on the magnitude of $\vec{b}$. We note that our use of the grain coordinate system for this definition is important because it is the system in which the crystallographic vectors are natively defined. Based on the expression in equation \ref{eq:deltaI}, the magnitude of $\varepsilon_0$ for the corresponding value of $\textbf{F}_{ij}^d$ (in the dislocation coordinate system) would take the form 
\begin{equation}
    F_{xx}^d = 1 - \frac{by}{4\pi(1-\nu)}\left[
    \frac{3x^2 + y^2 - 2\nu(x^2+y^2)}{(x^2+y^2)^2}
\right]
\end{equation}
\begin{equation}
F_{xy}^d = \frac{bx}{4\pi(1-\nu)}\left[
    \frac{3x^2 + y^2 - 2\nu(x^2+y^2)}{(x^2+y^2)^2}
\right]
\label{eq:1}
\end{equation}
\begin{equation}
F_{yx}^d = -\frac{bx}{4\pi(1-\nu)}\left[
    \frac{x^2 + 3y^2 - 2\nu(x^2+y^2)}{(x^2+y^2)^2}
\right]
\label{eq:2}
\end{equation}
\begin{equation}
F_{yy}^d = 1 + \frac{by}{4\pi(1-\nu)}\left[
    \frac{x^2 - y^2 - 2\nu(x^2+y^2)}{(x^2+y^2)^2}
\right]
\label{eq:3}
\end{equation}
\begin{equation}
F_{zz}^d = 1
\label{eq:4}
\end{equation}
that could be solved for based on the local contrast, $\delta I$, for a given point in the sample, $(x,y)$ in the dislocation system. We note that a transformation is required to convert the above equations into the grain system for comparison to our contrast function in Equation~\ref{eq:contrast}, where all the components will thus be nonzero. As the Poisson's coefficient is well established for most materials, we will use the most sensitive component of $\textbf{F}_{ij}^g$, i.e. $\textbf{F}_{zx}^{(g)}$ to find the $\vec{b}$ from a simulated DFXM image of a dislocation. $\textbf{F}_{ij}^g$ results in a simple equation

\begin{equation}
\textbf{F}_{zx}^g = 1 - \frac{bx(x^2 + 3y^2 - 2v(x^2 + y^2))}{4\pi(x^2 + y^2)^2(v - 1)} - \frac{by(3y^2 - x^2 + 2v(x^2 + y^2))}{4\pi(x^2 + y^2)^2(v - 1)}
\label{eqgrain}
\end{equation}

To obtain the Burgers vector precisely from $\delta I (\varepsilon_{0})$ expressions as in equation \ref{eq:deltaI}, we require 5 equations to solve for the 5 unknowns. This makes the process very cumbersome as we would need to simulate a $\phi$ and a $\chi$ scan at every $2\Delta\theta$ along 5 diffraction peaks within the range of the instrumental resolution function. We use the information from this section to derive the invisibility criterion for DFXM which is discussed in Section 6.

This approach gives us a seemingly easy way to measure $\vec{b}$, but is not viable for real experiments because this approach does not consider dynamical diffraction and cannot solve for the full vectorial components of $\vec{b}$ required to interpret the full character of the dislocation.

\section{Comparison of TEM and DFXM Invisibility Criteria}

The geometrical optics forward model for DFXM allows us to interrogate the existence of invisibility criteria from Section~\ref{sec: TEM theory of burgers vector identification}. We explore invisibility in this section by defining a region of interest that contains only one dislocation. To test the invisibility criterion in DFXM, we use the global definition of contrast to ensure our approach is invariant to the subtle changes caused by dynamical diffraction.

For the purposes of this work, we simulate an edge dislocation in FCC aluminum with line vector [112], burgers vector $[1\Bar{1}0]$ and slip plane normal $[11\Bar{1}]$. 
Following the measurement procedure outlined in Section~\ref{sec: TEM theory of burgers vector identification}, we model DFXM images of the dislocation along different diffraction vectors $\Vec{Q}$ to identifying which planes cause the dislocation to become invisible, satisfying the $\vec{Q}\cdot\vec{b} = 0$.
We select the $\langle 111 \rangle$ and $\langle 112 \rangle$ to test the TEM invisibility criteria. 
Building on the TEM theory, we sought $\vec{Q}_{hkl}$ vectors that would be mutually orthogonal to $\vec{b}$. For an edge dislocation, the slip plane normal, $\vec{n}$ and the line vector, $\vec{\ell}$, natively that satisfy this condition. As such, we first simulated our dislocation imaged along all symmetry equivalent versions of the slip plane normal, $\vec{Q}_{hkl}=[11\Bar{1}]_g$, and all symmetry equivalent versions of the line vector, $\vec{Q}_{hkl}=[112]_g$ for our simplified system.

We use the same synchrotron experimental setup shown in Dresselhaus-Marais et al. \cite{DresselhausMarais2021}, with the Simplified Geometry introduced in Section~\ref{sec:Defining DFXM sensitivity with the Forward Model}. 
These simulations use a resolution function compiled from a Monte-Carlo simulation (as done in \cite{Poulsen2021}) for goniometer angles  $\phi = \chi = 0$.

\begin{figure}
  \centering
  \includegraphics[width=\linewidth]{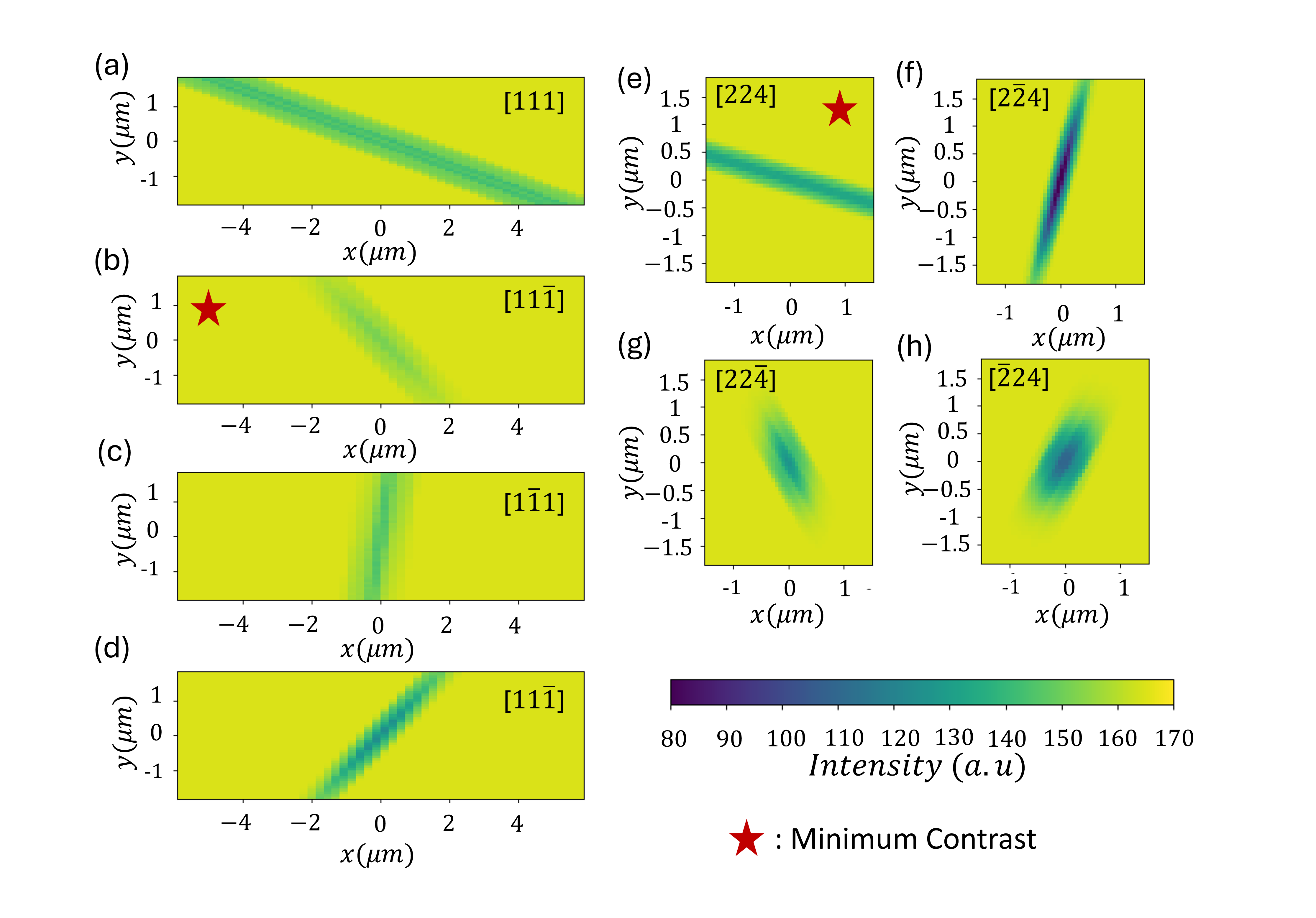}
  \caption{Contrast of a dislocation with $\Vec{n}_{s}=[11\Bar{1}]$, $\Vec{b}=[1\Bar{1}0]$, and $\Vec{\ell}=[112]$ in the strong beam condition imaged along different $\vec{Q}$ belonging to \{111\} and \{112\}; red star showing the images with minimum contrast.}
  \label{fig:strong beam dislocation contrast for 111}
\end{figure}

 In practical setups, CCD detector has a square using which the simulations have been performed. When we stretch this pattern into real space, the image appears as a rectangle rather than a square representing the actual physical dimensions of the area being studied.  A $2\theta$ angle of $45^\circ$ corresponds to the optimal condition for obtaining a perfect square simulated image. The closer the $2\theta$ angle approaches $45^\circ$, the more closely the simulated image resembles a square in real space.

Figure~\ref{fig:strong beam dislocation contrast for 111} show the images we simulated for all of the [111] diffraction vectors, and for 4 of the representative [211] diffraction vectors of the FCC lattice. 
As predicted by TEM, we observe strong contrast in $[\Bar{1}11]$ and $[1\Bar{1}2]$ which have the highest magnitude of $\vec{b}$ projected onto $\vec{Q}_{hkl}$. Similarly, we see a clear minimum contrast for $[11\Bar{1}]$ and $[112]$ both of which have zero projection of $\vec{b}$ onto $\vec{Q}_{hkl}$.
From the above results, we can see that the two minimum contrasts observed by DFXM produce a cross product that is equal to the Burgers vector. This validates that the TEM theory for dislocation invisibility is valid for DFXM experiments.

\begin{table}
    \centering
    \renewcommand{\arraystretch}{1.5} 

    \begin{tabular}{|c|c|c|c|}
        \hline
        \textbf{$\vec{Q}$} & \textbf{$I_{\text{max}}$ (arbitrary units)} & \textbf{$I_{\text{min}}$  (arbitrary units)} & \textbf{Global contrast ($I_{\text{max}}-I_{\text{min}}$)} \\ \hline
        [111] & 165.03 & 141.26 & 23.76 \\ \hline
        [11$\bar{1}$] & 165.03 & 151.91 & 13.11 \\ \hline
        [1$\bar{1}$1] & 164.09 & 144.54 & 19.54\\ \hline
        [$\bar{1}$11] & 165.03 & 123.88 & 41.14 \\       
        \hline
        [112] & 165.03 & 132.73 & 32.29 \\ \hline
        [1$\bar{1}$2] & 165.03 & 75.64 & 89.39 \\ \hline
        [11$\bar{2}$] & 165.03 & 126.11 & 38.92 \\ \hline
        [$\bar{1}$12] & 165.03 & 109.30 & 55.73 \\ \hline
    \end{tabular}
    \label{tab:my-table}
    \caption{Global contrast values for a dislocation with $\Vec{n}_{s}=[11\Bar{1}]$, $\Vec{b}=[1\Bar{1}0]$, and $\Vec{\ell}=[112]$ in the strong beam condition imaged along $\vec{Q}$ belonging to  \{111\} and \{112\}.}
   
\end{table}

Experimentally, however, we note some key differences between DF-TEM and DFXM measurements of dislocations. Though both electron and X-ray diffraction are governed by the same laws (e.g., von Laue conditions, extinction rules, etc.), the effective wavelengths of electron beams are typically much shorter than those used for X-ray diffraction. The shorter wavelengths produces an Ewald sphere of larger radius that intersect more points of the reciprocal lattice; this produces more reflections in electron diffraction than are typically observed for X-ray diffraction experiments. This results in electron microscopes having easier access to a wider range of diffraction peaks for DF-TEM measurements. By contrast, the DFXM of high-symmetry crystals like FCC lattice typically produce only a single diffraction peak in any given orientation for the 17-35 keV typically used.

Taking Al (111) diffraction as an example, the $\sim$100-300 keV energies typically used in TEMs give rise to $2\theta$ of $\sim 2-3^{\circ}$, as compared to the $2\theta=20.73^{\circ}$ angle generated by 17 keV X-ray beams used for DFXM. Because the angular range defining reciprocal space peaks is an order of magnitude broader by XRD, we note that DFXM is able to interrogate the deformation fields surrounding dislocations in higher deformation resolution than is possible by TEM.
We note, however, that while these simulations are relatively quick to perform, the experimental version of realigning the microscope to all 8 diffraction peaks would be a very time-consuming process ($\sim$8-hrs per peak).

\section{Extending DFXM to Imaging Dislocations in the Strong-Beam and Weak-Beam Conditions}

To identify an experimentally viable approach to measure $\vec{b}$, we now explore the contrast mechanism for a single-peak using DFXM scanning modalities. To understand how DFXM samples dislocations requires an understanding of the associated  deformation field (Equations~\ref{eq:1}-\ref{eq:4}) surrounding the dislocation cores. 
The deformation fields manifest in a broadening of diffraction peak profiles in reciprocal space. This 3D broadening of the $hkl$ peak arises from the tilt and strain fields of the crystal lattice that surround the dislocation core \cite{tang2007situ}.

DFXM utilizes extensive stacks of images acquired from angular scans along two different axes in the reciprocal space to map the long-range displacement fields around a dislocation core within each voxel of the sample. During these scans, the sample is either rocked through various $\phi$ positions (rocking scan) or $\chi$ positions (rolling scan) to observe symmetry offsets of these goniometer angles for a specific voxel to obtain strain and orientation information within the material.

\begin{equation}
   \frac{\Delta Q_{rock}}{|Q_0|}= -\frac{2\theta-2\theta_B}{2}- sin(\omega)(\chi-\chi_0) + cos(\chi)cos(\omega)(\phi-\phi_0)
   \label{eq:rock}
\end{equation}

\begin{equation}
   \frac{\Delta Q_{roll}}{|Q_0|}=  cos(\omega)(\chi-\chi_0) + cos(\chi)sin(\omega)(\phi-\phi_0)
   \label{eq:roll}
\end{equation}

We have observed that the TEM criteria $\vec{Q}\cdot(\vec{b}\times\vec{\ell})=0$ for invisibility is not sufficient to account for the appearance and disappearance of dislocations along a single ($hk\ell$) vector, $\vec{Q}_0$ across a 2 dimensional rocking curve ($\chi,\phi$) for DFXM.

\begin{figure}
  \centering
  \includegraphics[width=\linewidth]{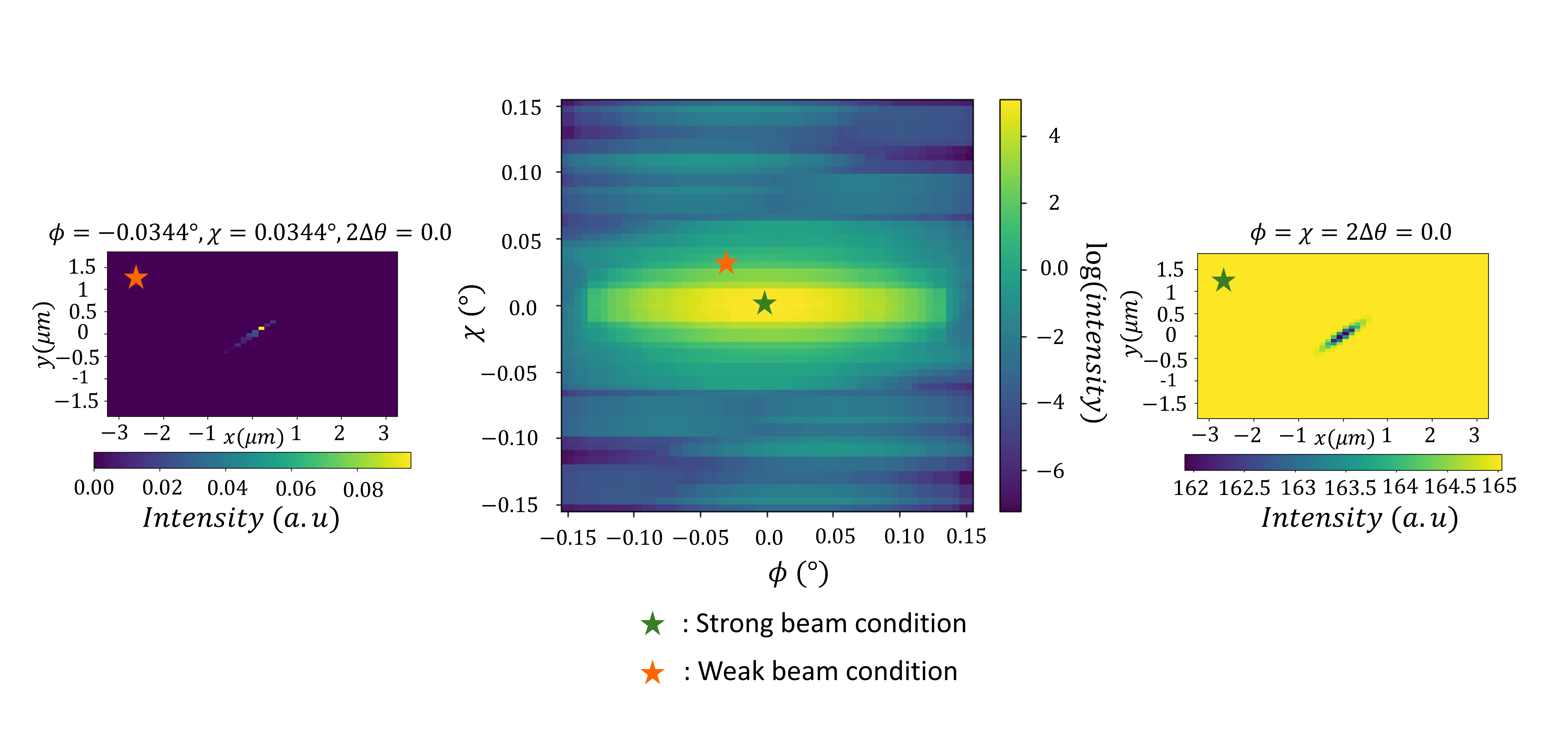}
  \caption{2D $\chi$-$\phi$ map of a dislocation with $\Vec{Q}=[002]$, $\Vec{n}_{s}=[11\Bar{1}]$, $\Vec{b}=[1\Bar{1}0]$, and $\Vec{\ell}=[112]$ with simulated DFXM images in the strong beam condition at position $\phi=\chi=2\Delta\theta=0^\circ$ and the weak beam condition at $\phi=-0.0344^\circ,\chi=0.0344^\circ, 2\Delta\theta=0^\circ$.}
  \label{fig:chi-phi-map}
\end{figure}

To demonstrate this effect, we simulated a characteristic two-dimensional $\phi$-$\chi$ scan in Figure \ref{fig:chi-phi-map} that would be performed experimentally, in the Simplified Geometry. We show each image as a pixel in a $\phi$ vs $\chi$ plot, with a color defining its magnitude of $\log{I_{\text{max}}}$ to show the variation along the rocking curve.

For two positions: $\phi=\chi=2\Delta\theta=0^\circ$ and $\phi=-0.0344^\circ,\chi=0.0344^\circ, 2\Delta\theta=0^\circ$, we plot the DFXM images showing two characteristic imaging conditions in Figure \ref{fig:chi-phi-map}: the \textit{strong-beam} condition when the crystal is in its pristine state and perfectly aligned to the diffraction condition, thereby revealing the full extent of its undistorted lattice structure, and the \textit{weak-beam} condition near a dislocation when the crystal becomes slightly distorted resulting in anomalous packing states \cite{yildirim2023extensive}.

\section{Identifying $\vec{b}$ in DFXM using the Weak Beam Condition}

DFXM signal in our simplified geometry includes contributions from only three components of the deformation gradient tensor.
These components include two shear deformations, $H_{13}^g$ and $H_{23}^g$, and one axial, $H_{33}^g$, given by
\begin{equation} \label{eq:qi}
    \Vec{q_i} \approx \begin{bmatrix}
    \cos(\theta_{0})H_{13}^g + \sin(\theta_{0})H_{33}^g \\
    H_{23}^g \\
    -\sin(\theta_{0})H_{13}^g + \cos(\theta_{0})H_{33}^g
    \end{bmatrix},
\end{equation}
as defined in \cite{Poulsen2021}, where $\Vec{q_i}$ is the normalised diffraction vector under consideration by DFXM, in the imaging coordinate system.
We use a normalized differential diffraction vector, $\Vec{q}$, in each coordinate system to define small offsets to the diffraction vectors under consideration, as DFXM probes a narrow range of reciprocal space. As such, the normalized differential $\vec{q}$ vectors are all referenced to the same undeformed $\vec{Q}_0$ by
\begin{equation}
    \Vec{q}= \frac{\vec{Q}-\vec{Q}_{0}}{|\vec{Q}_{0}|} .  
\end{equation}

To understand how to consider dislocation visibility by DFXM, we thus require an expression that defines the values of our goniometer angles that will generate zero contrast in the $H_{13}^g$, the $H_{23}^g$ and the $H_{33}^g$ components. This requires that we set the value of Equation \ref{eq:qi} equal to the vector $[0,0,0]^{T}$, which results in two conditions for DFXM invisibility in our simplified system,
\begin{align} 
    \tan{(\theta_{0})}=&-\frac{H_{13}}{H_{33}}, \text{    and} \\
    H_{23}^g=&0.
\end{align}

In TEM, the global image contrast produced by dislocations depends on the magnitude of $\vec{Q}_0\cdot\vec{b}$ for the selected $\vec{Q}_0$ vector. 
However, DFXM experiments have shown that the same dislocation line becomes visible and invisible across different rocking curve scans of the same nominal $\vec{Q}_0$ diffraction vector \cite{yildirim2023extensive}. Leveraging this observation, we now explore how different $\chi$-$\phi$ scans along a single $\vec{Q}_{0}$ can measure $\vec{b}$ directly. 

In dislocation science, $\vec{b}$ uniquely defines the shear imposed upon the lattice by a dislocation. As such, $\vec{b}$ defines a symmetry axis in reciprocal space that could uniquely define the native symmetry in $\bf{F}^{d}(x,y,z)$. 
A typical $\phi$ or $\chi$ scan measures the change in diffraction contrast along a vector of the reciprocal lattice, as defined in the sample coordinate system based on $\bf{\Gamma}$.
We thus predicted that for the appropriate crystallographic orientation, the goniometer angles defined in $\bf{\Gamma}$ that set the relationship between the sample and dislocation systems should dictate the symmetry of the dislocation images. 
As $\phi$ is the most sensitive scanning axis of DFXM's resolution function, if one were to perform a $\phi$ scan that samples along a vector that is close but misaligned to $\Vec{b}$, the images would sample deformation fields with that are asymmetric in $\bf{F}^{s}$, which is not aligned to the native symmetry of the dislocation system, $\bf{F}^{d}$. If the goniometer angles in $\bf{\Gamma}$ align the $\bf{F}^{d}$ and $\bf{F}^{s}$ systems such that a $\phi$ scan collects images along the $\vec{b}$ vector in reciprocal-space, it should produce images whose weak-beam contrast is symmetric in real-space images collected at identical values of $|\phi|$ about the strong-beam condition (for a fixed value of $\chi$ and $2\Delta\theta$). 

\begin{figure}
\centering
\includegraphics[width=\textwidth]{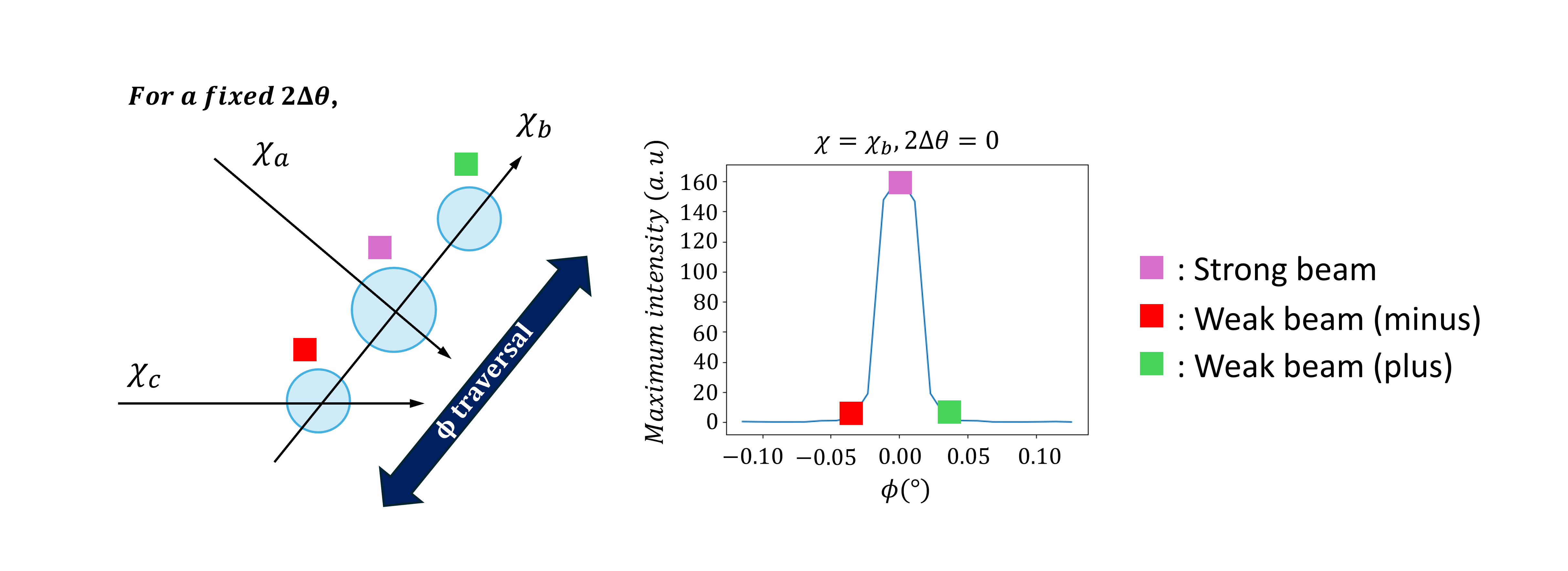}
\caption{Schematic of three $\chi$ rotations ($\chi_a$, $\chi_b$, $\chi_c$) sampling the reciprocal space in different ways at fixed $2\Delta\theta$ where $\chi_a$ slices through the strong beam, $\chi_b$ slices through only one  of the weak beams and $\chi_c$ samples the entire dislocation line. $\phi$ traversal at $\chi_b$ encodes information regarding the dislocation's burgers vector.}
\label{fig:schematic_hypothesis}
\end{figure}

To test how the alignment of $\Vec{b}$ with respect to the scanning axis gives rise to symmetrical or asymmetrical images along the rocking curve in DFXM, we here simulate images of the same edge dislocation considered above, for the diffraction peak $\vec{Q}_0 = [002]$ in the simplified geometry and measure the global contrast and look for symmetry of the dislocation features. We redefined the crystal orientation by defining $y_{c} = \Vec{b}$ to align our $\phi$ scan such that it simulates images that sample points aligned to the Burgers vector. If we orient $\Vec{Q}_0$ such that $\Vec{y}_{c}$ is set equal to $\Vec{b}$, the $\phi$ rotation thus rotates the sample system about $\vec{b}$ for all sampled values of $Q$. 

\begin{figure}
  \centering
  \includegraphics[width=\linewidth]{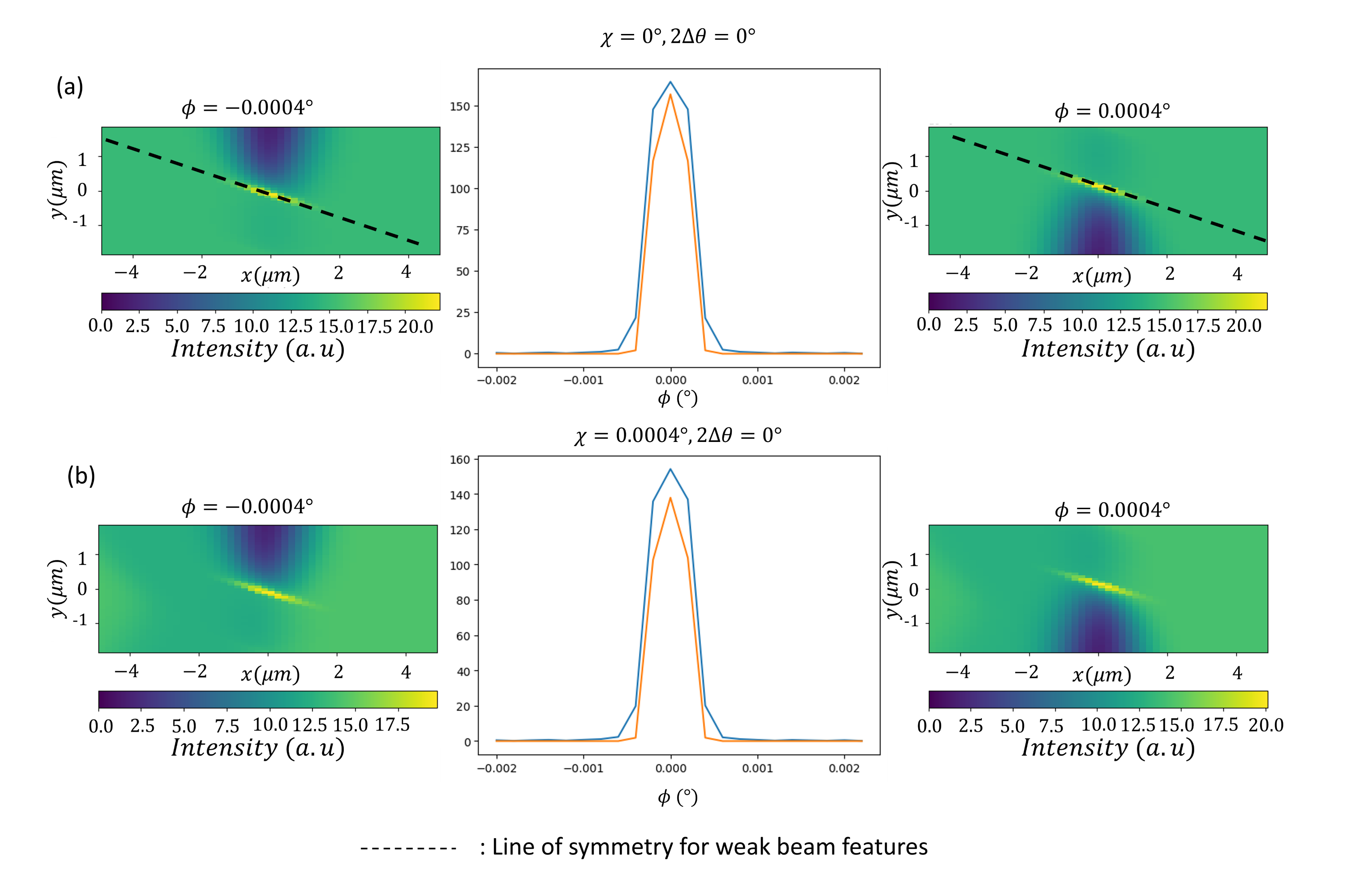}
  \caption{(a) Symmetry of weak beam images across a $\phi$ scan along $\vec{b}$ and (b) Asymmetry of weak beam images across a $\phi$ scan along a $\chi$ that is slightly offset in the positive direction with $\Delta\chi=0.0004^{\circ}$ of a dislocation with $\Vec{n}_{s}=[11\Bar{1}]$, $\Vec{b}=[1\Bar{1}0]$, and $\Vec{\ell}=[112]$ under the imaging conditions $hk\ell = [002]$, $\vec{x}_{\text{lab}} = [\bar{2}\bar{2}0]$, and $\vec{y}_{\text{lab}} = \vec{b}$.}
  \label{fig:weak beam dislocation contrast for 220}
\end{figure}

As shown in Fig.~\ref{fig:weak beam dislocation contrast for 220}a, $\phi$ is rocked from $-0.002^{\circ}$ to $0.002^{\circ}$ in very fine steps of $0.0002^{\circ}$, with unchanging values of $\chi = 0.000^{\circ}$ and $2\Delta \theta = 0$. We simulated 22 images, and plot the 1D rocking curve with the function $I_{\text{max}}(\phi)$ (blue)and the function $I_{\text{min}}(\phi)$ (orange). We note that while the 1D traces for the rocking-curve intensity are symmetric, this 1D measurement is insufficient to conclude alignment to $\vec{b}$ because the image features are most sensitive to asymmetry in $\bf{F}^s$. 

The images in Fig.~\ref{fig:weak beam dislocation contrast for 220} display the transition point between the strong- and weak-beam conditions, for values of $\phi=-0.0004^{\circ}$ and $\phi=0.0004^{\circ}$. As emphasized by the black dotted line, we observe image bright and dark image features that are symmetric in our simulated DFXM images when reflected along the Burgers vector. This confirms our initial predictions that dislocations that are visible with DFXM along a given $\vec{Q}_0$ diffraction peak do produce images whose features are symmetric along $\vec{b}$ when aligned to the Burgers vector.

To understand how this manifests for a \textit{misaligned} orientation, we show in Fig.~\ref{fig:weak beam dislocation contrast for 220}b, an analogous plot of a 1D rocking curve for the misaligned case in which $\phi$ does not scan along $\vec{b}$. In this case, we misalign the sample and dislocation systems by rotating the $\chi$ motor by $\Delta \chi=0.0004^{\circ}$. For this example, the $\phi$ scan is only slightly misaligned to $\vec{b}$, generating an asymmetric 1D rocking curve. By contrast, the images for the same $|\phi|$ values exhibit asymmetry in the dislocation's shape that is representative of the misalignment between coordinate systems.

\begin{figure}
  \centering
 \includegraphics[width=\linewidth]{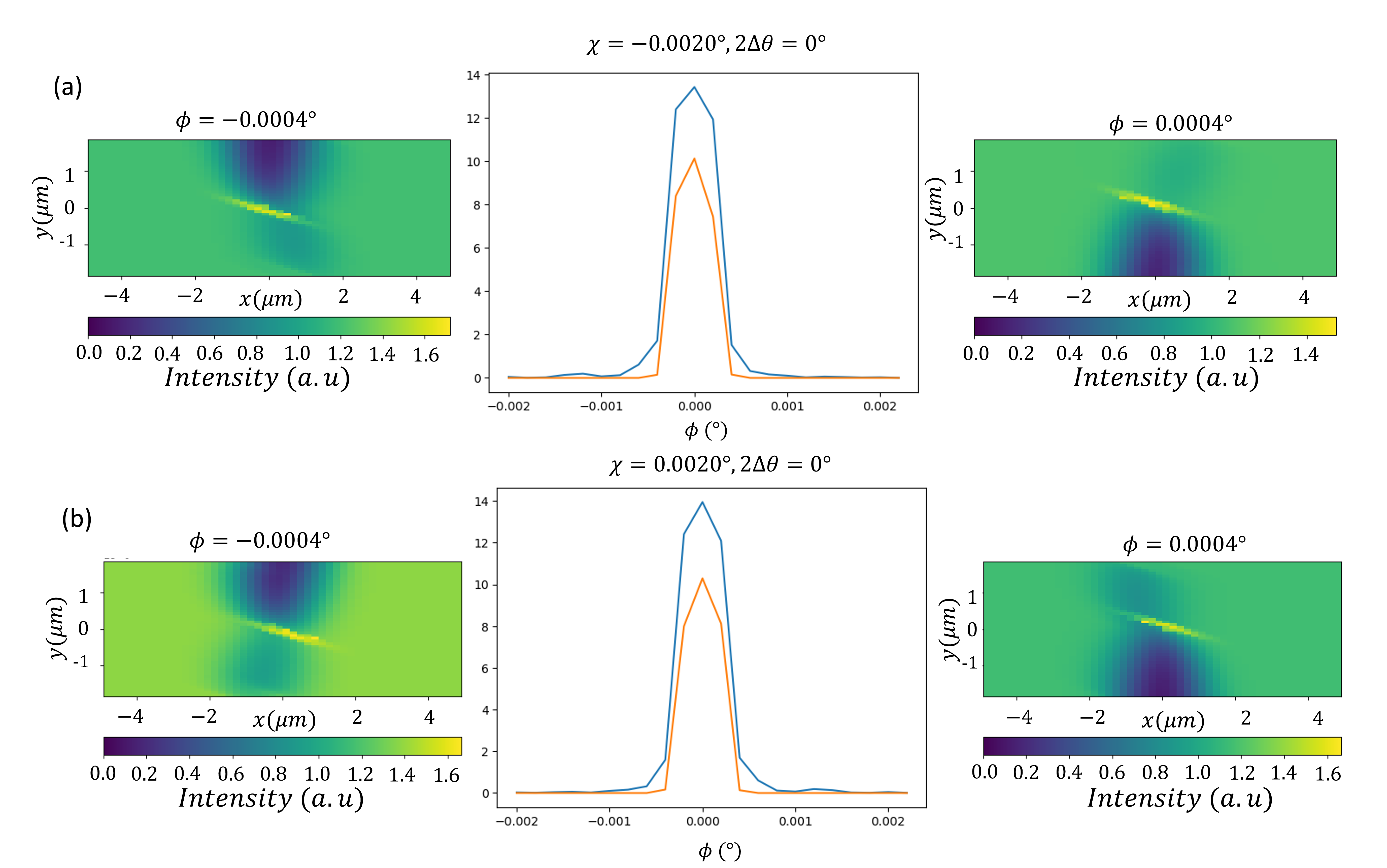}
  \caption{(a) Asymmetry of weak beam images across a $\phi$ scan along a $\chi$ that is offset in the negative direction where $\Delta\chi=-0.0020^{\circ}$ and (b) Asymmetry of weak beam images across a $\phi$ scan along a $\chi$ that is offset in the positive direction where $\Delta\chi=0.0020^{\circ}$ of a dislocation with $\Vec{n}_{s}=[11\Bar{1}]$, $\Vec{b}=[1\Bar{1}0]$, and $\Vec{\ell}=[112]$ under the imaging conditions $hk\ell = [002]$, $\vec{x}_{\text{lab}} = [\bar{2}\bar{2}0]$, and $\vec{y}_{\text{lab}} = \vec{b}$.}
    \label{fig:high_misalignment}
\end{figure}

Figure \ref{fig:high_misalignment} demonstrates an example with two cases where the $\chi$ value is way off with $\Delta \chi=0.0020^{\circ} and -0.0020^{\circ}$. The 1D rocking curve is asymmetric due to an offset in singularity and the asymmetric images on either side of the rocking curve show that even when the rounding error is insufficient to change our result, we would still not see the original symmetry from aligned $\vec{y_c}$ and $\vec{b}$.

With the four cases described above, we note that the quantitative offset between $\vec{y}_c$ and $\vec{b}$ in reciprocal space is essential to defining the feasibility of this measurement strategy for DFXM experiments. The simulations demonstrate that a $\chi$ offset of at least $0.0004^\circ$ is essential to resolve the asymmetry in the weak beam images. The step sizes in $\chi$ usually taken in real DFXM experiments is $0.004^\circ$ which is an order of magnitude higher than what this method requires to resolve $\vec{b}$. A similar step size of $\Delta\phi = 0.0004^\circ$ is ideal to be able to measure $\vec{b}$ without uncertainty caused by rounding errors \cite{Brennan2022}.

From the above examples, one may see that the symmetry of $\bf{F}^d$ and the positioning of $\bf{\Gamma}$ to define the relationship between $\vec{y}_c$ and $\vec{x}_d=\vec{b}$ offers a new strategy to directly measure $\vec{b}$ using DFXM along only one $\vec{Q}_0$ peak. This approach requires comparison in the symmetry of the image features that define the dislocations across a $\chi$-$\phi$ scan (i.e. the motors that sample \textit{shear} in the crystal).

\section{Discussion}

The difficulty in measurements described above arises in the limit of low signal-to-noise ratio when the transmission of the dark-field signal breaks down and due to rounding errors occuring from DFXM's high sensitivity to imprecision in motor positions.

In DFXM and in TEM, the detailed approaches to $\vec{b}$ identification is starting to be substituted for data science and computational approaches to interpret $\vec{b}$. For example, the image matching technique which compares the overall characteristics of theoretical and experimental images under various diffraction conditions for a specific dislocation \cite{head1967identification}. OVITO (Open Visualisation Tool) code \cite{stukowski2009visualization} with its built-in Dislocation Extraction Algorithm framework \cite{stukowski2012automated} uses Delaunay tesselation to identify dislocations and their burgers vector. We note that computational tools are being developed that seek to perform these Burgers vector identification problems in DFXM for FCC lattices. Our theoretical study offers the physical insights into those contrast mechanisms. This is important, as it offers direct measurement strategies for crystal systems that have a large number of possible Burgers vectors, as is seen in BCC crystals, or for systems in which $\vec{b}$ values appear that are unexpected and therefore not included in the training data.

\section{Conclusion}

Our theoretical study compares the TEM theory from Howie Whelan to the DFXM geometric optics model from Poulsen, et al. In this work, we demonstrate that the TEM models of $\Vec{Q}\cdot \Vec{b}=0$ do hold for DFXM, but are impractical for the constraints of DFXM experiments. We then display how the 2D rocking curves sampled by $\phi$ and $\chi$ offer direct sampling of the small distortions about the rocking curve that display anomalous deformation fields characteristic of the dislocations. Finally, we demonstrate how the dislocation symmetry may be aligned to the crystal system to directly measure $\vec{b}$ via the symmetry of the images collected along the rocking curve. Our work defines the first direct measurement of $\vec{b}$ to remove ambiguity for dislocation studies using DFXM, and offers the opportunity to now measure dislocation characters from DFXM images.

\appendix
\section{Invisibility- TEM and X- ray topography}

Dislocation invisibility, occurs when $(\Vec{Q} \cdot \Vec{b}) = 0$ for screw dislocations, while edge dislocations have an additional criteria of $\Vec{Q}\cdot (\Vec{b}\times \Vec{\ell}) = 0$, where $\Vec{Q}$ is the reciprocal lattice vector, $\Vec{b}$ is the burgers vector, and $\Vec{\ell}$ is the line vector of the dislocation. Since $\Vec{\ell}$ can be determined from the image and $\Vec{Q}$ can be determined by tilting the sample to achieve the invisibility criterion, then $\Vec{b}$ can be determined, which has proven to be a powerful application of dark-field electron microscopy in determining dislocation characters, structures and energies. In terms of the cylindrical polar coordinates, the displacement $R$  due to a mixed dislocation can be written as:
\begin{equation}
\vec{R} = \frac{1}{2\pi} \left[\vec{b}\theta + \vec{b_{e}} {\left(\frac{\sin 2\theta}{4\left(1 - \nu\right)}\right)} + (\vec{b}\times \vec{\ell})\{\left(\frac{\left(1 - 2\nu\right)}{2\left(1 -\nu\right)}\right){\ln} r + \frac{\cos 2\theta}{4\left(l -\nu\right)}\right\}]
\end{equation}

where $\Vec{b_{e}}$ is the edge component of $\Vec{b}$ \cite{head2012computed}.

\appendix
\section{Derivation of the Most General Case of TEM Invisibility Criteria}

The displacement field associated with a dislocation line is given by 
\begin{equation}
\Vec{u} = u_{x} \hat{i} + u_{y} \hat{j} + u_{z} \hat{k},
\end{equation}
where the components of $\Vec{u}$ can be defined as,
\begin{equation}
u_{x} = \frac{b_{e}}{2\pi} \left( \tan^{-1}\left(\frac{y}{x}\right) + \frac{xy}{2(1-\nu)(x^2+y^2)} \right)
\end{equation}

\begin{equation}
u_{y} = \frac{-b_{e}}{2\pi} \left( \frac{(1-2\nu)}{4(1-\nu)}\ln{(x^2+y^2)^2} + \frac{x^2-y^2}{4(1-\nu)(x^2+y^2)^2} \right)
\end{equation}

\begin{equation}
u_{z} = \frac{b_{s}}{2\pi} \tan^{-1}\left(\frac{y}{x}\right) 
\end{equation}

where $\Vec{b_{s}}$ is the screw component of $\Vec{b}$ \cite{Loethe1982}.

In diffraction based imaging of dislocations, it is understood that a dislocation line will become invisible when the projection of its displacement field on the diffraction vector $\vec{Q}$ is 0 \cite{gandais1982dislocation}.

\begin{equation}
\vec{Q} = q_{1} \hat{i} + q_{2} \hat{j} + q_{3} \hat{k}
\end{equation}

\begin{equation}
\vec{Q} \cdot \vec{u} = 0
\end{equation}

From this, we can write the most general form of the invisibility criterion for a dislocation line:

\begin{equation}
\begin{split}
q_{1} \cdot \left(\frac{\vec{b}_{e}}{2\pi} \left( \tan^{-1}\left(\frac{y}{x}\right) + \frac{xy}{2(1-\nu)(x^2+y^2)} \right)\right) + \\
q_{2} \cdot \left( \frac{-\vec{b}_{e}}{2\pi} \left( \frac{(1-2\nu)}{4(1-\nu)}\ln{(x^2+y^2)^2} + \frac{x^2-y^2}{4(1-\nu)(x^2+y^2)^2} \right)\right) + \\
q_{3} \cdot \left( \frac{\vec{b}_{s}}{2\pi} \tan^{-1}\left(\frac{y}{x}\right)\right) = 0
\end{split}
\end{equation}

\ack{Acknowledgements}

Funding for this research was sponsored by the Army Research Office and was accomplished under Cooperative Agreement Number W911NF-24-2-0085. The views and conclusions contained in this document are those of the authors and should not be interpreted as representing the official policies, either expressed or implied, of the Army Research Office or the U.S. Government. The U.S. Government is authorized to reproduce and distribute reprints for Government purposes notwithstanding any copyright notation herein.
Salary contributions for D.P. were supported by Stanford Graduate Fellowship, and for Y. W.'s by the Stanford Energy Fellows Postdoctoral program.

\referencelist{iucr}


@article{DresselhausMarais2021,
   author = {Leora E Dresselhaus-Marais and Grethe Winther and Marylesa Howard and Arnulfo Gonzalez and Sean R Breckling and Can Yildirim and Philip K Cook and Mustafacan Kutsal and Hugh Simons and Carsten Detlefs and Jon H Eggert and Henning Friis Poulsen},
   journal = {Sci. Adv},
   pages = {8311-8325},
   title = {In situ visualization of long-range defect interactions at the edge of melting},
   volume = {7},
   url = {https://www.science.org},
   year = {2021},
}

@article{Poulsen2017,
	author = {Poulsen, H. F. and Jakobsen, A. C. and Simons, H. and Ahl, S. R. and Cook, P. K. and Detlefs, C.},
	doi = {10.1107/S1600576717011037},
	journal = {Journal of Applied Crystallography},
	month = {Oct},
	number = {5},
	pages = {1441--1456},
	title = {{X-ray diffraction microscopy based on refractive optics}},
	url = {https://doi.org/10.1107/S1600576717011037},
	volume = {50},
	year = {2017}}

@article{Poulsen2021,
   author = {H. F. Poulsen and L. E. Dresselhaus-Marais and M. A. Carlsen and C. Detlefs and G. Winther},
   doi = {10.1107/s1600576721007287},
   issn = {1600-5767},
   issue = {6},
   journal = {Journal of Applied Crystallography},
   month = {12},
   pages = {1555-1571},
   publisher = {International Union of Crystallography (IUCr)},
   title = {Geometrical-optics formalism to model contrast in dark-field X-ray microscopy},
   volume = {54},
   year = {2021},
}

@article{Brennan2022,
  doi = {10.48550/ARXIV.2203.05671},
  
  url = {https://arxiv.org/abs/2203.05671},
  
  author = {Brennan, Michael C. and Howard, Marylesa and Marzouk, Youssef and Dresselhaus-Marais, Leora E.},
  
  title = {Analytical Methods for Superresolution Dislocation Identification in Dark-Field X-ray Microscopy},
  
  publisher = {arXiv},
  
  year = {2022},
  
  copyright = {Creative Commons Attribution 4.0 International}
}

@book{Loethe1982,
address = {New York},
author = {Hirth, John Price and Lothe, Jens},
edition = {2nd},
publisher = {Wiley},
title = {{Theory of Dislocations}},
year = {1982}
}

@article{Poulsen2018,
   author = {H. F. Poulsen and P. K. Cook and H. Leemreize and A. F. Pedersen and C. Yildirim and M. Kutsal and A. C. Jakobsen and J. X. Trujillo and J. Ormstrup and C. Detlefs},
   doi = {10.1107/S1600576718011378},
   issn = {16005767},
   issue = {5},
   journal = {Journal of Applied Crystallography},
   month = {10},
   pages = {1428-1436},
   publisher = {Wiley-Blackwell},
   title = {Reciprocal space mapping and strain scanning using X-ray diffraction microscopy},
   volume = {51},
   year = {2018},
}

@article{chen2020atomic,
  title={Atomic Force Microscopy Study of Discrete Dislocation Pile-ups at Grain Boundaries in Bi-Crystalline Micro-Pillars},
  author={Chen, Xiaolei and Richeton, Thiebaud and Motz, Christian and Berbenni, St{\'e}phane},
  journal={Crystals},
  volume={10},
  number={5},
  pages={411},
  year={2020},
  publisher={MDPI}
}

@article{kuhlmann1999theory,
  title={The theory of dislocation-based crystal plasticity},
  author={Kuhlmann-Wilsdorf, D},
  journal={Philosophical Magazine A},
  volume={79},
  number={4},
  pages={955--1008},
  year={1999},
  publisher={Taylor \& Francis}
}

@book{head2012computed,
  title={Computed electron micrographs and defect identification},
  author={Head, AK},
  volume={7},
  year={2012},
  publisher={Elsevier}
}

@article{gandais1982dislocation,
  title={Dislocation contrast by transmission electron microscopy A method for Burgers vector characterization if the invisibility criterion is not valid},
  author={Gandais, M and Hihi, A and Willaime, C and Efelboin, Y},
  journal={Philosophical Magazine A},
  volume={45},
  number={3},
  pages={387--400},
  year={1982},
  publisher={Taylor \& Francis}
}

@article{stinville2016sub,
  title={Sub-grain scale digital image correlation by electron microscopy for polycrystalline materials during elastic and plastic deformation},
  author={Stinville, JC and Echlin, MP and Texier, Damien and Bridier, F and Bocher, P and Pollock, TM},
  journal={Experimental mechanics},
  volume={56},
  pages={197--216},
  year={2016},
  publisher={Springer}
}

@article{taheri2016current,
  title={Current status and future directions for in situ transmission electron microscopy},
  author={Taheri, Mitra L and Stach, Eric A and Arslan, Ilke and Crozier, Peter A and Kabius, Bernd C and LaGrange, Thomas and Minor, Andrew M and Takeda, Seiji and Tanase, Mihaela and Wagner, Jakob B and others},
  journal={Ultramicroscopy},
  volume={170},
  pages={86--95},
  year={2016},
  publisher={Elsevier}
}

@article{hirsch1956lxviii,
  title={LXVIII. Direct observations of the arrangement and motion of dislocations in aluminium},
  author={Hirsch, PB and Horne, RW and Whelan, MJ},
  journal={Philosophical Magazine},
  volume={1},
  number={7},
  pages={677--684},
  year={1956},
  publisher={Taylor \& Francis}
}

@article{yildirim2023extensive,
  title={Extensive 3D mapping of dislocation structures in bulk aluminum},
  author={Yildirim, Can and Poulsen, Henning F and Winther, Grethe and Detlefs, Carsten and Huang, Pin H and Dresselhaus-Marais, Leora E},
  journal={Scientific Reports},
  volume={13},
  number={1},
  pages={3834},
  year={2023},
  publisher={Nature Publishing Group UK London}
}

@article{lang1958direct,
  title={Direct observation of individual dislocations by x-ray diffraction},
  author={Lang, AR},
  journal={Journal of Applied Physics},
  volume={29},
  number={3},
  pages={597--598},
  year={1958},
  publisher={American Institute of Physics}
}

@article{danilewsky2020x,
  title={X-Ray Topography—More than Nice Pictures},
  author={Danilewsky, Andreas N},
  journal={Crystal Research and Technology},
  volume={55},
  number={9},
  pages={2000012},
  year={2020},
  publisher={Wiley Online Library}
}

@article{hammarberg2023strain,
  title={Strain Mapping of Single Nanowires using Nano X-ray Diffraction},
  author={Hammarberg, Susanna},
  year={2023}
}

@article{clark2015three,
  title={Three-dimensional imaging of dislocation propagation during crystal growth and dissolution},
  author={Clark, Jesse N and Ihli, Johannes and Schenk, Anna S and Kim, Yi-Yeoun and Kulak, Alexander N and Campbell, James M and Nisbet, Gareth and Meldrum, Fiona C and Robinson, Ian K},
  journal={Nature materials},
  volume={14},
  number={8},
  pages={780--784},
  year={2015},
  publisher={Nature Publishing Group UK London}
}

@article{simons2015dark,
  title={Dark-field X-ray microscopy for multiscale structural characterization},
  author={Simons, Hugh and King, A and Ludwig, Wolfgang and Detlefs, C and Pantleon, Wolfgang and Schmidt, S{\o}ren and St{\"o}hr, F and Snigireva, I and Snigirev, A and Poulsen, Henning Friis},
  journal={Nature communications},
  volume={6},
  number={1},
  pages={6098},
  year={2015},
  publisher={Nature Publishing Group UK London}
}

@article{jakobsen2019mapping,
  title={Mapping of individual dislocations with dark-field X-ray microscopy},
  author={Jakobsen, AC and Simons, H and Ludwig, W and Yildirim, C and Leemreize, H and Porz, L and Detlefs, C and Poulsen, HF},
  journal={Journal of Applied Crystallography},
  volume={52},
  number={1},
  pages={122--132},
  year={2019},
  publisher={International Union of Crystallography}
}

@article{borgi2024simulations,
  title={Simulations of dislocation contrast in dark-field X-ray microscopy},
  author={Borgi, Sina and R{\ae}der, Trygve Magnus and Carlsen, Mads Allerup and Detlefs, Carsten and Winther, Grethe and Poulsen, Henning Friis},
  journal={Journal of Applied Crystallography},
  volume={57},
  number={2},
  year={2024},
  publisher={International Union of Crystallography}
}

@article{bernier2011far,
  title={Far-field high-energy diffraction microscopy: a tool for intergranular orientation and strain analysis},
  author={Bernier, Joel Vincent and Barton, Nathan Rhodes and Lienert, Ulrich and Miller, Matthew Peter},
  journal={The Journal of Strain Analysis for Engineering Design},
  volume={46},
  number={7},
  pages={527--547},
  year={2011},
  publisher={SAGE Publications Sage UK: London, England}
}

@article{head1967identification,
  title={The identification of burgers vectors and unstable directions of dislocations in $\beta$-brass},
  author={Head, AK and Loretto, MH and Humble, P},
  journal={physica status solidi (b)},
  volume={20},
  number={2},
  pages={521--536},
  year={1967},
  publisher={Wiley Online Library}
}

@article{stukowski2009visualization,
  title={Visualization and analysis of atomistic simulation data with OVITO--the Open Visualization Tool},
  author={Stukowski, Alexander},
  journal={Modelling and simulation in materials science and engineering},
  volume={18},
  number={1},
  pages={015012},
  year={2009},
  publisher={IOP Publishing}
}

@article{stukowski2012automated,
  title={Automated identification and indexing of dislocations in crystal interfaces},
  author={Stukowski, Alexander and Bulatov, Vasily V and Arsenlis, Athanasios},
  journal={Modelling and Simulation in Materials Science and Engineering},
  volume={20},
  number={8},
  pages={085007},
  year={2012},
  publisher={IOP Publishing}
}

@article{tang2007situ,
  title={An in situ method for the study of strain broadening using synchrotron X-ray diffraction},
  author={Tang, Chiu C and Lynch, Peter A and Cheary, Robert W and Clark, Simon M},
  journal={Journal of Applied Crystallography},
  volume={40},
  number={4},
  pages={642--649},
  year={2007},
  publisher={International Union of Crystallography}
}
\end{document}